\documentclass[12pt]{article}
\usepackage{amsmath,amsfonts,amsthm,amssymb,eucal}
\newcommand{\Qp}{\mathbb Q_p}
\newcommand{\DA}{D^\alpha}
\newcommand{\DAN}{D^\alpha_N}
\newcommand{\DD}{\mathcal D(\Qp)}
\newcommand{\DDD}{\mathcal D'(\Qp)}

\newcommand{\FF}{\mathcal F}
\numberwithin{equation}{section}
\begin{document}
\newtheorem{lem}{Lemma}
\newtheorem*{teo}{Theorem}
\newtheorem{prop}{Proposition}
\newtheorem*{defin}{Definition}
\pagestyle{plain}
\title{On the $p$-Adic Navier-Stokes Equation}
\author{ \textbf{Andrei Yu. Khrennikov}\\
\footnotesize International Center for Mathematical Modeling\\
\footnotesize in Physics and Cognitive Sciences,\\
\footnotesize Mathematical Institute, Linnaeus University,\\
\footnotesize V\"axj\"o, SE-35195 Sweden,\\
\footnotesize E-mail: andrei.khrennikov@lnu.se
\and
\textbf{Anatoly N. Kochubei}\\
\footnotesize Institute of Mathematics,\\
\footnotesize National Academy of Sciences of Ukraine,\\
\footnotesize Tereshchenkivska 3, Kiev, 01004 Ukraine,\\
\footnotesize E-mail: kochubei@imath.kiev.ua }

\date{}
\maketitle

\bigskip
\begin{abstract}
We prove the local solvability of the $p$-adic analog of the Navier-Stokes equation. This equation describes, within the $p$-adic model of porous medium, the flow of a fluid in capillaries.
\end{abstract}
\vspace{2cm}


\newpage

\section{Introduction}

A $p$-adic model of propagation of fluids through the capillary structure of a porous medium was suggested in \cite{KOC1}. In this model an idealized fragment of a porous medium is identified with the $p$-adic ball interpreted as the set of (generally infinite) paths of a homogeneous rooted tree of valence $p+1$. Here $p$ is a fixed prime number.

Natural developments prompted by this idea include both new mathematical models of percolation phenomena \cite{KOC2,KOC3,AOKK} and purely mathematical works dealing with $p$-adic analogs of equations of mathematical hydrodynamics, such as the porous medium equation \cite{KK,K2018}.

The $p$-adic Navier-Stokes equation \cite{KOC3} is a pseudo-differential evolution equation on the field $\Qp$ of $p$-adic numbers describing average velocity of a fluid moving through the $p$-tree of capillaries. This nonlinear equation (deduced in \cite{KOC3} from the disretized model of hydrodynamics \cite{BBTT}) has the form
\begin{equation}
\label{1.1}
\frac{\partial u(t,x)}{\partial t}=u(t,x)(D^1 u)(t,x)-\theta (D^2u)(t,x)
\end{equation}
where $\theta >0$, $D^\alpha$ ($\alpha >0$) is the Vladimirov fractional differentiation operator on $\Qp$ (see \cite{K2001,VVZ}).

In this paper we initiate the mathematical theory of the equation (\ref{1.1}). Note that there exists a well-developed theory of linear pseudo-differential equations on $\Qp$; see \cite{AKS,KKZ,K2001,Zu}. The study of nonlinear equations of this kind is only beginning \cite{KK,K2018}.

Our method of investigating the equation (\ref{1.1}) is based on abstract results by von Wahl \cite{Wa} who found sufficient conditions of local solvability of the Cauchy problem for the equation
\begin{equation}
\label{1.2}
v'(t)+Av(t)+M(v(t))=0,\quad t>0,
\end{equation}
where $A$ is the generator of an analytic semigroup $e^{-tA}$ in a Banach space $\mathfrak B$, $M$ is a nonlinear operator subordinated to $A^{1-\rho}$, $0<\rho <1$; see the precise formulation in Section 2 below.

In our situation, where we study the local solvability of (\ref{1.1}) on a bounded domain, $\mathfrak B=L^q(B_N)$ where $1<q<\infty$, $B_N=\left\{ x\in \Qp:\ |x|_p\le p^N\right\}$ is a $p$-adic ball, the operators $A$ and $M$ are constructed from the Vladimirov operators $D^2$ and $D^1$ on the ball. Note that the latter operators are nonlocal, which makes even the definition of an operator on a bounded domain nontrivial; see \cite{K2001,K2018}. In addition, there is an $L^2$-theory of the Vladimirov operator \cite{AKS,KKZ,K2001,VVZ} and the initial steps towards its $L^1$-theory \cite{KK}. Here we have to develop its $L^q$-theory, in particular to prove certain inequalities for various $L^q$-norms involving $D^\alpha u$ and $D^\beta u$, $0<\alpha <\beta$.

The structure of this paper is as follows. In Section 2 we give main notions regarding $p$-adic numbers and the Vladimirov operators, as well as an introduction to von Wahl's theory. Section 3 is devoted to the $L^q$-theory of the operators $D^\alpha$, while in Section 4 we prove the inequalities mentioned above. Finally, in Section 5 we formulate and prove our result on the local solvability of the Cauchy problem for the equation (\ref{1.1}) on a $p$-adic ball.

\medskip
\section{Preliminaries}

{\bf 2.1.} {\it $p$-Adic numbers} \cite{VVZ}. Let $p$ be a prime
number. The field of $p$-adic numbers is the completion $\mathbb Q_p$ of the field $\mathbb Q$
of rational numbers, with respect to the absolute value $|x|_p$
defined by setting $|0|_p=0$,
$$
|x|_p=p^{-\nu }\ \mbox{if }x=p^\nu \frac{m}n,
$$
where $\nu ,m,n\in \mathbb Z$, and $m,n$ are prime to $p$. $\Qp$ is a locally compact topological field. By Ostrowski's theorem there are no absolute values on $\mathbb Q$, which are not equivalent to the ``Euclidean'' one, or one of $|\cdot |_p$.

The absolute value $|x|_p$, $x\in \mathbb Q_p$, has the following properties:
\begin{gather*}
|x|_p=0\ \mbox{if and only if }x=0;\\
|xy|_p=|x|_p\cdot |y|_p;\\
|x+y|_p\le \max (|x|_p,|y|_p).
\end{gather*}

The latter property called the ultra-metric inequality (or the non-Archi\-me\-dean property) implies the total disconnectedness of $\Qp$ in the topology
determined by the metric $|x-y|_p$, as well as many unusual geometric properties. Note also the following consequence of the ultra-metric inequality: $|x+y|_p=\max (|x|_p,|y|_p)$, if $|x|_p\ne |y|_p$.

The absolute value $|x|_p$ takes the discrete set of nonzero
values $p^N$, $N\in \mathbb Z$. If $|x|_p=p^N$, then $x$ admits a
(unique) canonical representation
\begin{equation}
\label{2.1}
x=p^{-N}\left( x_0+x_1p+x_2p^2+\cdots \right) ,
\end{equation}
where $x_0,x_1,x_2,\ldots \in \{ 0,1,\ldots ,p-1\}$, $x_0\ne 0$.
The series converges in the topology of $\mathbb Q_p$. For
example,
$$
-1=(p-1)+(p-1)p+(p-1)p^2+\cdots ,\quad |-1|_p=1.
$$
We denote $\mathbb Z_p=\{ x\in \Qp:\ |x|_p\le 1\}$. $\mathbb Z_p$, as well as all balls in $\Qp$, is simultaneously open and closed.

Proceeding from the canonical representation (\ref{2.1}) of an element $x\in
\mathbb Q_p$, one can define the fractional part of $x$ as the rational number
$$
\{ x\}_p=\begin{cases}
0,& \text{if $N\le 0$ or $x=0$};\\
p^{-N}\left( x_0+x_1p+\cdots +x_{N-1}p^{N-1}\right) ,& \text{if
$N>0$}.\end{cases}
$$
The function $\chi (x)=\exp (2\pi i\{ x\}_p)$ is an additive
character of the field $\mathbb Q_p$, that is a character of its additive group. It is clear
that $\chi (x)=1$ if and only if $|x|_p\le 1$.

Denote by $dx$ the Haar measure on the
additive group of $\Qp $ normalized by the equality $\int_{\mathbb Z_p}dx=1$.

The additive group of $\Qp$ is self-dual, so that
the Fourier transform of a complex-valued function $f\in L^1(\Qp)$ is again a function on $\Qp$ defined as
$$
(\mathcal Ff)(\xi )=\int\limits_{\Qp}\chi (x\xi )f(x)\,dx.
$$
If $\mathcal Ff\in L^1(\Qp)$, then we have the inversion formula
$$
f(x)=\int\limits_{\Qp}\chi (-x\xi )\widetilde{f}(\xi )\,d\xi .
$$
It is possible to extend $\mathcal F$ from $L^1(\Qp)\cap L^2(\Qp)$ to a unitary operator on $L^2(\Qp)$, so that the Plancherel identity holds in this case.

In order to define distributions on $\Qp$, we have to specify a class of test functions. A function $f:\ \Qp\to \mathbb C$ is called locally constant if
there exists such an integer $l\ge 0$ that for any $x\in \Qp$
$$
f(x+x')=f(x)\quad \mbox{if }\|x'\|\le p^{-l}.
$$
The smallest number $l$ with this property is called the exponent of local constancy of the function $f$.

Typical examples of locally constant functions are additive characters, and also cutoff functions like
$$
\Omega (x)=\begin{cases}
1,& \text{if $\|x\|\le 1$};\\
0,& \text{if $\|x\|>1$}.\end{cases}
$$
In particular, $\Omega$ is continuous, which is an expression of the non-Archimedean properties of $\Qp$.

Denote by $\mathcal D(\Qp)$ the vector space of all locally constant functions with compact supports. Note that $\DD$ is dense in $L^q(\Qp)$ for each $q\in [1,\infty )$. In order to furnish $\DD$ with a topology, consider first the subspace $D_N^l\subset \DD$ consisting of functions with supports in a ball
$$
B_N=\{ x\in \Qp:\ |x|_p\le p^N\},\quad N\in \mathbb Z,
$$
and the exponents of local constancy $\le l$. This space is finite-dimensional and possesses a natural direct product topology. Then the topology in $\DD$ is defined as the double inductive limit topology, so that
$$
\DD=\varinjlim\limits_{N\to \infty}\varinjlim\limits_{l\to \infty}D_N^l.
$$

If $V\subset \Qp$ is an open set, the space $\mathcal D(V)$ of test functions on $V$ is defined as a subspace of $\DD$ consisting of functions with supports in $V$. For a ball $V=B_N$, we can identify $\mathcal D(B_N)$ with the set of all locally constant functions on $B_N$. The set $\mathcal D(B_N)$ is dense in $L^q(B_N)$, $1\le q<\infty$.

The space $\DDD$ of Bruhat-Schwartz distributions on $\Qp$ is defined as a strong conjugate space to $\DD$.

\medskip
{\bf 2.2.} {\it Vladimirov's operator} \cite{AKS,K2001,VVZ}.

The Vladimirov operator $\DA$, $\alpha >0$, of fractional differentiation, is defined first as a pseudo-differential operator with the symbol $|\xi|_p^\alpha$:
\begin{equation}
\label{2.2}
(\DA u)(x)=\FF^{-1}_{\xi \to x}\left[ |\xi |_p^{\alpha }\FF_{y\to \xi }u\right] ,\quad u\in \DD,
\end{equation}
where we show arguments of functions and their direct/inverse Fourier transforms. There is also a hypersingular integral representation giving the same result on $\DD$ but making sense on much wider classes of functions (for example, bounded locally constant functions):
\begin{equation}
\label{2.3}
\left( \DA u\right) (x)=\frac{1-p^\alpha }{1-p^{-\alpha -1}}\int\limits_{\Qp}|y|_p^{-\alpha -1}[u(x-y)-u(x)]\,dy.
\end{equation}

The Cauchy problem for the heat-like equation
$$
\frac{\partial u}{\partial t}+\DA u=0,\quad u(0,x)=\psi (x),\quad x\in\Qp,t>0,
$$
is a model example for the theory of $p$-adic parabolic equations. If $\psi$ is regular enough, for example, $\psi \in \DD$, then a classical solution is given by the formula
$$
u(t,x)=\int\limits_{\Qp}Z_\alpha(t,x-\xi )\psi (\xi )\,d\xi
$$
where $Z_\alpha$ is, for each $t$, a probability density and
$$
Z_\alpha(t_1+t_2,x)=\int\limits_{\Qp}Z_\alpha(t_1,x-y)Z_\alpha (t_2,y)\,dy,\quad t_1,t_2>0,\ x\in \Qp.
$$

The "heat kernel" $Z_\alpha$ can be written as the Fourier transform
\begin{equation}
\label{2.4}
Z_\alpha(t,x)=\int\limits_{\Qp}\chi (\xi x)e^{-t|\xi |_p^\alpha}\,d\xi .
\end{equation}
See \cite{K2001} for various series representations and estimates of the kernel $Z_\alpha$.

The natural stochastic process in $B_N$ corresponds to the Cauchy problem
\begin{equation}
\frac{\partial u(t,x)}{\partial t}+\left( \DAN u\right) (t,x)-\lambda_\alpha u(t,x)=0,\quad x\in B_N,t>0;
\end{equation}
\begin{equation}
u(0,x)=\psi (x),\quad x\in B_N,
\end{equation}
where $\lambda_\alpha =\dfrac{p-1}{p^{\alpha +1}-1}p^{\alpha (1-N)}$, the operator $\DAN$ is defined by restricting $\DA$ to functions $u_N$ supported in $B_N$ and considering the resulting function $\DA u_N$ only on $B_N$. Note that $\DAN$ defines a positive definite selfadjoint operator on $L^2(B_N)$, $\lambda$ is its smallest eigenvalue.

Under certain regularity assumptions, for example if $\psi \in \mathcal D(B_N)$, the problem (2.5)-(2.6) possesses a classical solution
$$
u(t,x)=\int\limits_{B_N}Z_N^{(\alpha )}(t,x-y)\psi (y)\,dy,\quad t>0,x\in B_N,
$$
where $Z_N^{(\alpha )}(t,x)\ge 0$,
\begin{equation}
\label{2.7}
Z_N^{(\alpha )}(t,x)=e^{\lambda_\alpha t}Z_\alpha (t,x)+c(t),
\end{equation}
$$
c(t)=p^{-N}-p^{-N}(1-p^{-1})e^{\lambda_\alpha t}\sum \limits _{n=0}^\infty
\frac{(-1)^n}{n!}t^n\frac{p^{-N\alpha n}}{1-p^{-\alpha n-1}}.
$$
The kernel $Z_N^{(\alpha )}$ satisfies the identity
\begin{equation}
\label{2.8.1}
\int\limits_{B_N}Z_N^{(\alpha )}(t,x)\,dx=1.
\end{equation}

It was shown in \cite{KK} that the family of operators
\begin{equation}
\label{2.8}
(T_N^{(\alpha)}(t)\psi )(x)=\int\limits_{B_N}Z_N^{(\alpha)}(t,x-y)\psi (y)\,dy
\end{equation}
is a strongly continuous contraction semigroup on $L^1(B_N)$. Its generator $A_N^{(\alpha )}$ coincides with $\DAN -\lambda_\alpha I$ at least on $\mathcal D(B_N)$. More generally, this is true in the distribution sense on restrictions to $B_N$ of functions from the domain of the generator of the semigroup on $L^1(\Qp )$ corresponding to $\DA$.

Another approach to the operator $\DAN$ and its extensions was developed in \cite{K2018} where it was interpreted as a pseudo-differential operator, in terms of the Pontryagin duality on $B_N$ ($B_N$ is an additive group, due to the ultrametric inequality).

\medskip
{\bf 2.3.} {\it A theorem by von Wahl}. Let us consider the equation (\ref{1.2}) with the initial condition $v(0)=\varphi$, $\varphi \in D(A)$, where a linear operator $A$ is the generator of an analytic semigroup in a Banach space $\mathfrak B$, $M$ is a nonlinear operator in $\mathfrak B$. It is assumed that, for some $\rho \in (0,1)$, $M$ acts from the domain $D(A^{1-\rho})$ to $\mathfrak B$ and satisfies the following condition: if $v,w\in D(A)$, $\|Av\|+\|Aw\|\le h$, $h>0$, then there exists such a constant $k(h)>0$ that
\begin{equation}
\label{2.9}
\begin{cases}
\|M(u)-M(v)\| \le k(h)\|A^{1-\rho}(v-w)\|,\\
\| M(v)\| \le k(h).
\end{cases}
\end{equation}
Here $\|\cdot \|$ means the norm in $\mathfrak B$.

The von Wahl theorem (\cite{Wa}, Theorem II.1.1) states that, under the above conditions, there exists $T(\varphi)\in (0,\infty]$ with the following property. For any $T<T(\varphi)$, there exists a unique function $v\in C^1([0,T],\mathfrak B)$, such that $Av\in C([0,T],\mathfrak B)$, satisfying the equation (\ref{1.2}) and the initial condition $v(0)=\varphi$.

\medskip
\section{The Vladimirov operator in $L^q(B_N)$, $q\ge 1$.}

Let $T_N^{(\alpha )}(t)$ be the family of operator defined by (\ref{2.8}).

\begin{prop}
For any $q\ge 1$, the family $T_N^{(\alpha )}$ is a $C_0$-semigroup of contraction operators on $L^q(B_N)$.
\end{prop}

\medskip
{\it Proof}. The family $T_N^{(\alpha )}$ is a positivity preserving Markov semigroup on $L^2(B_N)$ generated by a nonnegative selfadjoint operator $\DAN -\lambda_\alpha$ (see \cite{K2001}). On the other hand, it is a contraction $C_0$-semigroup on $L^1(B_N)$ \cite{KK}. By (\ref{2.8.1}), $T_N^{(\alpha )}(t)$ is a contraction on $L^\infty (B_N)$ for each $t$. It follows from these properties (\cite{Da}, Theorems 1.3.3 and 1.4.1) that $T_N^{(\alpha )}$ is a $C_0$-contraction semigroup on $L^q(B_N)$, $1\le q <\infty$. $\qquad \blacksquare$

\medskip
{\it Remark}. If $1<q<\infty$, then $T_N^{(\alpha )}$ is is holomorphic on the sector $|\arg z|<\frac{\pi}2 (1-|\frac2{q}-1|)$. See \cite{Da}, Theorem 1.4.2.

\medskip
Define a $m$-accretive linear operator $A_q^{(\alpha )}$ on $L^q(B_N)$ as a generator of the $C_0$-semigroup $T_N^{(\alpha )}$ on $L^q(B_N)$. Properties of the similar operator on $L^1(B_N)$ proved in \cite{K2018} (representation formulas for this operator and its resolvent, interpretation in the distribution sense) remain valid for the operator $A_q^{(\alpha )}$.

Below we understand $\DAN$ as an operator on $L^q(B_N)$, such that $A_q^{(\alpha )}=\DAN -\lambda_\alpha$. On $\mathcal D(B_N)$, $\DAN$ coincides with the operator defined in Section 2.2 in terms of the restriction procedure.

\medskip
\begin{prop}
$\mathcal D(B_N)$ is dense in the domain $D(A_q^{(\alpha )})$ with respect to the graph norm of the operator $A_q^{(\alpha )}$.
\end{prop}

\medskip
{\it Proof}. If $u\in D(A_q^{(\alpha )})$, set $f=\left( (A_q^{(\alpha )}+\mu I\right)u$, $\mu >0$. Let us approximate $f$ in $L^q(B_N)$ by a sequence $\{ f_m\}\subset \mathcal D(B_N)$ and set $u_m=(A_q^{(\alpha )}+\mu I)^{-1}f_m$. It follows from the convolution representation of the resolvent (see Theorem 2 in \cite{K2018}) that $u_m\in \mathcal D(B_N)$.

Obviously, $u_m\to (A_q^{(\alpha )}+\mu I)^{-1}f=u$ in $L^q(B_N)$ and
$$
A_q^{(\alpha )}u_m=(A_q^{(\alpha )}+\mu I)u_m-\mu u_m=f_m-\mu u_m\to f-\mu u=A_q^{(\alpha )}u
$$
in $L^q(B_N)$, as desired. $\qquad \blacksquare$

This property shows that a priori estimates involving $A_q^{(\alpha )}$ can be proved by checking the inequalities on functions from $\mathcal D(B_N)$.

\medskip
\section{Inequalities}

{\bf 4.1.} {\it Fractional powers}. Since $T_N^{(\beta)}$, $\beta >0$, is a contraction semigroup in $L^q(B_N)$, the operator $D_N^\beta$ is a generator of a semigroup of type $-\lambda_\beta$. For the definition of its fractional powers see $\S 5$, Chapter 1, in \cite{Kr}, in particular Theorem 5.6.

\medskip
\begin{prop}
Let $0<\alpha \le \beta$. For all $u\in \mathcal D(B_N)$,
\begin{equation}
\label{4.1}
\left\| D_N^\alpha u\right\|_{L^q(B_N)}\le C\left\| \left( D_N^\beta\right)^{\alpha/\beta} u\right\|_{L^q(B_N)}
\end{equation}
where $C$ does not depend on $u$ (here and below $C$ denotes various positive constants).
\end{prop}

\medskip
{\it Proof}. Following \cite{Kr} we use the representation
\begin{equation}
\label{4.*}
\left( D_N^\beta\right)^{-\alpha/\beta}f=\frac1{\Gamma \left(\frac{\alpha}{\beta}\right) }\int\limits_0^\infty t^{\frac{\alpha}{\beta}-1}e^{-\lambda_\beta t}T_N^{(\beta )}(t)f\,dt, \quad f\in L^q(B_N),
\end{equation}
where $e^{-\lambda_\beta t}T_N^{(\beta )}(t)$ is the convolution operator in $L^q(B_N)$ (recall that $B_N$ is an additive group) with the kernel
\begin{equation}
\label{4.2}
G_N^{(\beta )}(t,x)=Z_\beta (t,x)+e^{-\lambda_\beta t}c_\beta (t)
\end{equation}
where
\begin{equation}
\label{4.3}
c_\beta (t)=p^{-N}-e^{\lambda_\beta t}p^{-N}\int\limits_{B_N}Z_\beta (t,y)\,dy
\end{equation}
(see \cite{K2018}, especially the proof of Theorem 2).

We find from (\ref{4.2}) and (\ref{4.3}) that
\begin{equation}
\label{4.4}
G_N^{(\beta )}(t,x)=Z_\beta (t,x)-p^{-N}\int\limits_{B_N}Z_\beta (t,y)\,dy+p^{-N}e^{-\lambda_\beta t}.
\end{equation}
Next, using (\ref{2.4}) we have
$$
p^{-N}\int\limits_{B_N}Z_\beta (t,y)\,dy=p^{-N}\int\limits_{\Qp}e^{-t|\xi |_p^\beta}\,d\xi \int\limits_{B_N}\chi (-y\xi )\,dy=\int\limits_{|\xi|_p\le p^{-N}}e^{-t|\xi |_p^\beta}\,d\xi
$$
by a well-known integration formula \cite{VVZ}.

Let $x\in B_N$, $|x|_p=p^m$, $m\le N$. The same integration formula gives
$$
\int\limits_{|\xi |_p\ge p^{-m+2}}\chi (-x\xi )e^{-t|\xi |_p^\beta}\,d\xi =\sum\limits_{k=-m+2}^\infty e^{-tp^{k\beta}}\int\limits_{|\xi|_p=p^k}\chi (-x\xi )\,d\xi =0,
$$
so that
\begin{equation}
\label{4.5}
G_N^{(\beta )}(t,x)=\int\limits_{p^{-N+1}\le |\xi |_p\le p^{-m+1}}\chi (-x\xi )e^{-t|\xi |_p^\beta}\,d\xi +p^{-N}e^{-\lambda_\beta t}.
\end{equation}

Denote $\left( D_N^\beta\right)^{\alpha/\beta} u=f$. The desired inequality (\ref{4.1}) is equivalent to the inequality
\begin{equation}
\label{4.6}
\left\| \DAN \left( D_N^\beta\right)^{-\alpha/\beta}f\right\|_{L^q(B_N)}\le C\|f\|_{L^q(B_N)}.
\end{equation}
It follows from (\ref{4.*}) and (\ref{4.6}) that the convolution kernel of the operator $\left( D_N^\beta\right)^{-\alpha/\beta}$ equals
\begin{multline*}
\frac1{\Gamma \left(\frac{\alpha}{\beta}\right) }\int\limits_{p^{-N+1}\le |\xi |_p\le p^{-m+1}}\chi (-x\xi )d\xi \int\limits_0^\infty t^{\frac{\alpha}{\beta}-1}e^{-t|\xi|_p^\beta}\,dt+p^{-N}\frac1{\Gamma \left(\frac{\alpha}{\beta}\right) }\int\limits_0^\infty t^{\frac{\alpha}{\beta}-1}e^{-\lambda_\beta t}\,dt\\
=\int\limits_{p^{-N+1}\le |\xi |_p\le p^{-m+1}}\chi (-x\xi )|\xi|_p^{-\alpha}d\xi +p^{-N}\lambda_\beta^{-\frac{\alpha}\beta},
\end{multline*}
that is, by Theorem 2 from \cite{K2018}, it is equal to the convolution kernel of the operator $\left( \DAN\right) ^{-1}$ plus a constant $d=p^{-N}\left( \lambda_\beta^{-\frac{\alpha}\beta}-\lambda_\alpha^{-1}\right)$, so that
$$
\left( \left( D_N^\beta\right)^{-\alpha/\beta}f\right) (x)=\left( \left( \DAN\right)^{-1}f\right) (x)+d\int\limits_{B_N}f(y)\,dy.
$$

Now, from the integral representation of the operator $\DAN$ given in \cite{K2018} (Theorem 1, (ii)), it follows that
$$
\left( \DAN \left( D_N^\beta\right)^{-\alpha/\beta}f\right) (x)=f(x)+\lambda_\alpha d\int\limits_{B_N}f(y)\,dy,\quad x\in B_N,
$$
which implies (\ref{4.6}). $\qquad \blacksquare$

\medskip
{\bf 4.2.} {\it Comparison of fractional powers}. Let $0<\alpha <\beta \le \alpha +1$, $\beta >1$, $1\le q \le r<\infty$.

\medskip
\begin{prop}
Suppose that
\begin{equation}
\label{4.7}
\alpha -\beta +\frac1q <\frac1r \le \frac1q.
\end{equation}
Then for any $u\in \mathcal D(B_N)$
\begin{equation}
\label{4.8}
\left\| \DAN u\right\|_{L^r(B_N)}\le C\left\| D_N^\beta u\right\|_{L^q(B_N)}
\end{equation}
where the constant $C$ does not depend on $u$.
\end{prop}

\medskip
{\it Proof}. Let $f=D_N^\beta u$. Then $f\in \mathcal D(B_N)$ (see \cite{K2018}, Theorem 1), and the inequality (\ref{4.8}) is equivalent to the inequality
\begin{equation}
\label{4.9}
\left\| \DAN \left( D_N^\beta\right)^{-1}f\right\|_{L^r(B_N)}\le C\| f\|_{L^q(B_N)}.
\end{equation}

By Theorem 2 from \cite{K2018},
\begin{equation}
\label{4.**}
\left( \left( D_N^\beta\right)^{-1}f\right) (x)=\int\limits_{B_N}K_{\lambda_\beta}(x-\xi )f(\xi )\,d\xi +\lambda_\beta^{-1}p^{-N}\int\limits_{B_N}f(\xi )\,d\xi
\end{equation}
where
\begin{equation}
\label{4.10}
K_{\lambda_\beta}(x)=\int\limits_{|\lambda|_p\ge p^{-N+1}}\frac{\chi (\eta x)}{|\eta |_p^\beta}\,d\eta.
\end{equation}

Another result from \cite{K2018} (Theorem 1) is the representation
\begin{equation}
\label{4.11}
\left( \DAN \varphi \right) (x)=\lambda_\alpha \varphi (x)+\frac{1-p^\alpha}{1-p^{-\alpha -1}}\int\limits_{B_N}|y|_p^{-\alpha -1}[\varphi (x-y)-\varphi (x)]\,dy,\quad \varphi \in \mathcal D(B_N).
\end{equation}

Substituting (\ref{4.10}) into (\ref{4.11}) we see that
\begin{multline}
\label{4.12}
\left( \DAN \left( D_N^\beta\right)^{-1}f\right) (x)=\lambda_\alpha \int\limits_{B_N}K_{\lambda_\beta}(x-\xi )f(\xi )\,d\xi+\lambda_\alpha \lambda_\beta^{-1}p^{-N}\int\limits_{B_N}f(\xi )\,d\xi \\
+\frac{1-p^\alpha}{1-p^{-\alpha -1}}\int\limits_{B_N}|y|_p^{-\alpha -1}dy\int\limits_{B_N}\left[ K_{\lambda_\beta}(x-y-\xi )-K_{\lambda_\beta}(x-\xi )\right] f(\xi )\,d\xi.
\end{multline}

Note that
$$
K_{\lambda_\beta}(z-y)-K_{\lambda_\beta}(z)=\int\limits_{|\eta|_p\ge p^{-N+1}}|\eta|_p^\beta \chi (\eta z)[\chi (-\eta y)-1]\,d\eta,
$$
so that
\begin{multline*}
\frac{1-p^\alpha}{1-p^{-\alpha -1}}\int\limits_{B_N}|y|_p^{-\alpha -1}\left[ K_{\lambda_\beta}(z-y)-K_{\lambda_\beta}(z)\right] \,dy\\
=\frac{1-p^\alpha}{1-p^{-\alpha -1}}\int\limits_{|\eta|_p\ge p^{-N+1}}|\eta|_p^\beta \chi (\eta z)\,d\eta\int\limits_{B_N}|y|_p^{-\alpha -1}[\chi (-\eta y)-1]\,dy.
\end{multline*}

The internal integral in $y$ is known (see \cite{K2001}, page 164) -- if $|\eta|_p\ge p^{-N+1}$, then
$$
\int\limits_{B_N}|y|_p^{-\alpha -1}[\chi (-\eta y)-1]\,dy=|\eta|_p^\alpha -\lambda_\alpha,
$$
and we obtain the equality
\begin{multline}
\label{4.13}
\frac{1-p^\alpha}{1-p^{-\alpha -1}}\int\limits_{B_N}|y|_p^{-\alpha -1}\,dy\int\limits_{B_N}\left[ K_{\lambda_\beta}(x-y-\xi )-K_{\lambda_\beta}(x-\xi )\right] f(\xi )\,d\xi \\
=\int\limits_{B_N}f(\xi)\,d\xi \left[ \int\limits_{|\eta|_p\ge p^{-N+1}}|\eta|_p^{\alpha -\beta} \chi (\eta (x-\xi))\,d\eta-\lambda_\alpha \int\limits_{|\eta|_p\ge p^{-N+1}}|\eta|_p^{-\beta} \chi (\eta (x-\xi))\,d\eta \right]
\end{multline}
where the integrals in $\eta$ converge in the sense of $\mathcal D'(\Qp)$, so that the right hand side makes sense, since $f\in \mathcal D(B_N)$.

Now we substitute (\ref{4.13}) into (\ref{4.12}) and notice that the second integral in the right-hand side of (\ref{4.13}) cancels with the first summand in (\ref{4.12}). We have
\begin{equation}
\label{4.14}
\left( \DAN \left( D_N^\beta\right)^{-1}f\right) (x)=\lambda_\alpha \lambda_\beta^{-1}p^{-N}\int\limits_{B_N}f(\xi )\,d\xi +\int\limits_{B_N}I(x-\xi )f(\xi)\,d\xi
\end{equation}
where
$$
I(z)=\int\limits_{|\eta|_p\ge p^{-N+1}}|\eta|_p^\beta \chi (\eta z)\,d\eta,\quad z\in B_N.
$$

While this integral a priori exists in $\mathcal D'$, we can study its pointwise behavior using the representation
\begin{equation}
\label{4.15}
I(z)=\sum\limits_{n=-N}^\infty p^{n(\alpha -\beta)}\int\limits_{|\eta|_p=p^n}\chi (\eta z)\,d\eta.
\end{equation}
If $|z|_p=p^j$, $-\infty <j\le N$, then \cite{K2001,VVZ}
$$
\int\limits_{|\eta|_p=p^n}\chi (\eta z)\,d\eta=\begin{cases}
(1-\frac1p)p^n, & \text{if $n\le -j$};\\
-p^{n-1}, & \text{if $n=-j+1$};\\
0, & \text{if $n\ge -j+2$}.\end{cases}
$$
For $j=N$, only one term ($n=-N+1$) in (\ref{4.15}) is different from zero, and we get
$$
I(z)=-p^{(-N+1)(\alpha -\beta )}p^{-N}=-p^{\alpha -\beta}|z|_p^{-(\alpha -\beta +1)}.
$$

If $j<N$, then
$$
I(z)=(1-\frac1p)\sum\limits_{n=-N+1}^{-j}p^{n(\alpha -\beta +1)}-p^{(-j+1)(\alpha -\beta )}p^{-j}=(1-\frac1p)\sum\limits_{\nu =1}^{N-j}p^{(\nu -N)(\alpha -\beta +1)}-\frac1p\cdot p^{(-j+1)(\alpha -\beta +1)}.
$$
Computing the progression we find that in both cases
\begin{equation}
\label{4.16}
|I(z)|\le C|z|_p^{-(\alpha -\beta +1)},\quad z\in B_N.
\end{equation}

Let uas apply to the convolution on $B_N$, appearing in (\ref{4.14}), the Young inequality
$$
\|I*f\|_{L^r(B_N)}\le \|I\|_{L^s(B_N)}\|f\|_{L^q(B_N)}
$$
where $s,q,r\ge 1$, $s^{-1}+q^{-1}=1+r^{-1}$ (in \cite{We}, this inequality is proved for general groups, which covers the case of $B_N$). By (\ref{4.16}), the function $I$ belongs to $L^s(B_N)$, if $1\le s<\dfrac1{\alpha -\beta +1}$. In particular, if we choose $s$ from the equality $s^{-1}+q^{-1}=1+r^{-1}$, then by our assumption (\ref{4.7}) we have $1\le s<\dfrac1{\alpha -\beta +1}$.

Applying the H\"older inequality to the first summand in (\ref{4.14}) we come to the inequality (\ref{4.9}), which implies the required inequality (\ref{4.8}). $\qquad \blacksquare$

\medskip
\section{Local solvability of the $p$-adic Navier-Stokes equation}

Let us consider the equation
\begin{equation}
\label{5.1}
\frac{\partial u(t,x)}{\partial t}=u(t,x)\left(D_N^1 u\right) (t,x)-\theta \left(D_N^2 u\right) (t,x)
\end{equation}
with the initial condition
\begin{equation}
\label{5.2}
u(0,x)=\varphi (x).
\end{equation}
We apply von Wahl's theorem with $\mathfrak B=L^q(B_N)$, $1<q<\infty$, $A=\theta D_N^2$, $M(u)=u\cdot D_N^1u$. As we know, $A$ is a generator of an analytic subgroup in $\mathfrak B$.

\medskip
\begin{teo}
For any $\varphi \in D(A)$ (in particular, for any $\varphi \in \mathcal D(B_N)$), there exists $T(\varphi)\in (0,\infty]$ with the following property. For any $T$, $0<T<T(\varphi)$, the Cauchy problem (\ref{5.1}), (\ref{5.2}) possesses a unique solution $u\in C^1([0,T],\mathfrak B)$, such that $Au\in C([0,T],\mathfrak B)$ .
\end{teo}

\medskip
{\it Proof}. We have to check the conditions of von Wahl's theorem. Let $v,w\in D(A)$, $\|Av\|_{L^q(B_N)}+\|Aw\|_{L^q(B_N)}\le h$. By the H\"older inequality,
$$
\|M(v)\|_{L^q(B_N)}\le \|v\|_{L^{qr}(B_N)}\|D_N^1v\|_{L^{qs}(B_N)},\quad v\in \mathcal D(B_N),
$$
where $r,s\ge 1$, $\dfrac1r+\dfrac1s=1$.

By Proposition 4,
\begin{equation}
\label{5.3}
\|D_N^1v\|_{L^{qs}(B_N)}\le C \|\theta D_N^2v\|_{L^q(B_N)}, \quad v\in \mathcal D(B_N),
\end{equation}

It follows (with the use of the H\"older inequality) from (\ref{4.**}) and the continuity of the resolvent kernel for $D_N^2$ that
$$
\left\| \left(D_N^2\right)^{-1}f\right\|_{L^{qr}(B_N)}\le C\|f\|_{L^q(B_N)}, \quad f\in L^q(B_N),
$$
so that
$$
\|v\|_{L^{qr}(B_N)}\le  C \left\|\theta D_N^2v\right\|_{L^q(B_N)},\quad v\in D(A),
$$
and the second inequality in (\ref{2.9}) has been proved.

Next, $M(v)-M(w)=(v-w)D_N^1v+w\left(D_N^1v-D_N^1w\right)$. In subsequent estimates we can use $\left\|D_N^{2(1-\rho)}v\right\|_{L^q(B_N)}$ instead of $\|A^{(1-\rho)}v\|_{L^q(B_N)}$. That is possible by virtue of Proposition 3.

As before, we have for $0<\rho <\frac12$ that
$$
\left\|(v-w)D_N^1v\right\|_{L^q(B_N)}\le \|v-w\|_{L^{qr}(B_N)}\left\|D_N^1v\right\|_{L^{qs}(B_N)}\le C\left\|D_N^{2(1-\rho)}(v-w)\right\|_{L^q(B_N)}\left\|D_N^2v\right\|_{L^q(B_N)}.
$$

Finally, if $\rho <\dfrac12-\dfrac12(\dfrac1q-\dfrac1{qs})=\dfrac12(1-\dfrac1{qr})$, then by Proposition 4,
\begin{multline*}
\left\| w\cdot D_N^1(v-w)\right\|_{L^q(B_N)}\le \|w\|_{L^{qr}(B_N)}\left\|D_N^1(v-w)\right\|_{L^{qs}(B_N)}\\
\le C\left\| D_N^2w\right\|_{L^q(B_N)}\left\|D_N^{2(1-\rho)}(v-w)\right\|_{L^q(B_N)},
\end{multline*}
which proves the first inequality in (\ref{2.9}). $\qquad \blacksquare$

\section*{Acknowledgments}
The second author is grateful to Mathematical Institute, Linnaeus University, for hospitality during his visits to V\"axj\"o. The work of the second author was also supported in part by Grant 23/16-18 ``Statistical dynamics, generalized Fokker-Planck equations, and their applications in the theory of complex systems'' of the Ministry of Education and Science of Ukraine.

\end{document}